\documentclass[twocolumn,preprintnumbers,amsmath,amssymb,dvipdfmx]{revtex4-1}
\usepackage{graphicx}% Include figure files
\usepackage{dcolumn}% Align table columns on decimal point
\usepackage{bm}% bold math
\usepackage{color}
\usepackage{ulem}

\begin{document}

\title{
Spectroscopy of Majorana modes of non-Abelian vortices in Kitaev's chiral spin liquid
}% Force line breaks with \\

\author{Masafumi Udagawa$^{1,2}$ and Roderich Moessner$^2$}%
\affiliation{%
$^1$Department of Physics, Gakushuin University, Mejiro, Toshima-ku, Tokyo 171-8588, Japan\\
$^2$Max-Planck-Institut f\"{u}r Physik komplexer Systeme, 01187 Dresden, Germany
}%

\date{\today}% It is always \today, today,

\begin{abstract}
We study the temperature ($T$) dependence of the dynamical structure factor  of the chiral spin liquid phase of Kitaev's honeycomb model. 
We find, using a recently developed analytical approach, direct signatures 
of the Majorana modes associated with the non-Abelian $\mathbb{Z}_2$ 
vortices (visons). In particular, a characteristic multiplet of discrete peaks at {\it{low but nonzero-$T$}} and its field dependence reflect
 the existence of thermally 
activated visons whose varying separation yields different splittings of the Majorana `zero' modes. 
The resonant processes involved are specific to the Majorana modes, and will persist even in the presence of weak non-integrable interactions.
\end{abstract}

%\pacs{71.10.Fd, 71.10.Hf, 71.20.-b, 71.23.-k}% PACS, the Physics and Astronomy
%\keywords{Suested keywords}%Use showkeys class option if keyword
\maketitle
Majorana zero modes (MZM) represent particularly striking manifestation of the non-locality of quantum mechanics. 
They have been a focus of research for their promise of providing topological protection to quantum information by
encoding it non-locally, `fractionalised' involving two spatially separated objects \cite{Kitaev_2001}. 
Several systems have been proposed for their realization, ranging from one-dimensional $p$-wave superconductors~\cite{Kitaev_2001}, via non-abelian fractional quantum Hall systems~\cite{PhysRevB.61.10267} to the chiral spin liquid (CSL) phase of Kitaev's honeycomb model~\cite{kitaev2006anyons}. 

A most remarkable recent experiment on the quasi--two-dimensional magnet $\alpha$-RuCl$_3$ in a magnetic field
has found a half-integer quantization of thermal Hall conductivity~\cite{kasahara2018majorana} as expected for
a chiral Kitaev spin liquid \cite{kitaev2006anyons,PhysRevLett.121.147201,PhysRevX.8.031032} where the thermal current is carried
by a topologically protected edge state. Such a  gapped magnetic phase is also known to host 
an emergent $\mathbb{Z}_2$ gauge
field, the vortices of which are associated with a Majorana degree of freedom. In this work, we consider the question
if, and how, these degrees of freedom can be detected via bulk probes such as neutron scattering, which
has already been carried out in some detail on the zero-field phase ~\cite{banerjee2016proximate,banerjee2017neutron,2017NatPh..13.1079D}. A characteristic signature of the Majorana modes would be particularly desirable as it would in turn
amount to evidence for the existence of the non-Abelian vortices which host them, and the chiral spin liquid
in turn underpinning the existence of the vortices. All three conclusions would be important milestones for the field
of interacting topological phases generally, and quantum spin liquids in particular~\cite{Knolle1804.02037}. 

Spectroscopic experiments on $\alpha$-RuCl$_3$ in a field have advanced rapidly, via inelastic neutron scattering~\cite{Banerjeefield}, Raman spectroscopy~\cite{wulferding2019magnon}, electron spin resonance~\cite{PhysRevB.96.241107} and THz spectroscopy~\cite{PhysRevLett.119.227202}. All these studies found a complicated reconstruction of magnetic modes along with the breakdown of the magnetic phase. In particular, the former two studies claim to have found a characteristic evolution of resonant peaks with decreasing temperatures and associated it with Majorana excitations.

To complement these experimental developments, we here provide 
 a controlled theoretical analysis of the dynamics of Kitaev's chiral QSL. 
 The dynamical structure factor~\cite{knolle2014dynamics,PhysRevB.92.115127,PhysRevB.96.024438,samarakoon2018classical,yoshitake2017temperature,suzuki2018effective,yamaji2016clues,PhysRevLett.119.157203,PhysRevB.97.075126,PhysRevB.98.014418}, which we compute, provides information on the excitation spectrum both 
 of the ground state and, at nonzero temperatures, of the excited states. The latter aspect is important as 
 $T$ naturally controls the vison fugacity of the emergent gauge field, and hence provides access
 to a varying density of vison excitations. We are for the first time able to provide the full $T$-dependence 
 of the chiral spin liquid dynamics, {including low-$T$ regime beyond the reach of quantum Monte Carlo technique~\cite{yoshitake2019majoranamagnon},} thanks to the recently-developed analytical scheme~\cite{PhysRevB.98.220404} 
 which we here adapt
 to the chiral phase; this 
gives  us access to the real-time correlation function directly, obviating the need for an analytical continuation.

{In detail, we present a set of resonant peaks which appear in a  transient low-$T$ regime, below the  high-$T$ incoherent regime. Here, a low density of thermally excited visons carry Majorana zero modes which interact with each other. The members of the multiplet of peaks reflects different processes: the displacement of free visons; pair-creation of visons; and dissociation of vison pairs. Their differing physical origins lead to distinct temperature and magnetic field dependences of the peak height, as well as the resonant frequencies. These processes are characteristic of Majorana many-body physics and as such should be robust to the addition of non-integrable interactions.}

%---------%
\begin{figure}[h]
\begin{center}
\includegraphics[width=0.9\linewidth]{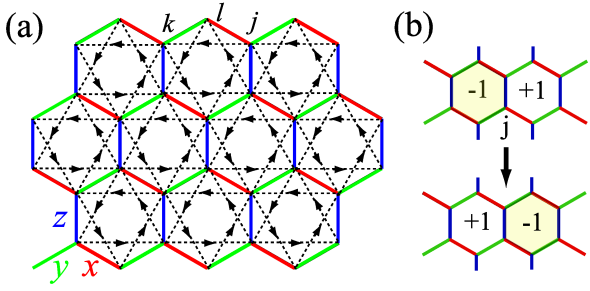}
\vspace{-10pt}
\caption{(Color online) (a) Honeycomb lattice. Three groups of bonds, $x$, $y$ and $z$ are colored in red, green and blue, respectively. A triplet of sites, $\langle j,k,l\rangle_{xy}$ is shown as an example. Other triplets are obtained by translations and combined spin and $\pi/3$-spatial rotations.
Dashed arrows point the phase of second-neighbor hopping in Eq.~(\ref{eq:MajoranaHamiltonian}), and it reverses if $\kappa$ takes a different sign.
(b) An example of $\mathbb{Z}_2$ fluxes, and the process of their change due to the operation of $S_{j}^z$.}
\label{fig1}
\end{center}
\end{figure}
%---------% 

{\it Model:} We consider the Kitaev's honeycomb model in a magnetic field, treating magnetic field in a perturbative way as pioneered in Ref.~\cite{kitaev2006anyons}, to obtain a soluble Hamiltonian for the CSL phase:
\begin{eqnarray}
\mathcal{H}=-J_{\rm K}\sum_{\langle i,j\rangle_{\alpha}}S_i^{\alpha}S^{\alpha}_{j} - 2\kappa\sum_{\langle j,k,l\rangle_{\alpha\beta}}S^{\alpha}_{j}S^{\beta}_{k}S^{\gamma}_{l}.
\label{eq:Hamiltonian}
\end{eqnarray}
Here, $S_i^{\alpha}$ is the $\alpha(=x,y,z)$ component of spin-$1/2$ operators defined on a honeycomb lattice. 
For the first term, we classify the bonds into three groups, $\langle i,j\rangle_{\alpha}$, along which the spins are coupled with Ising interactions of the assigned orientations [Fig.~\ref{fig1} (a)]. The second, three-spin interaction, term accounts for the magnetic field. An example of the triplets, $\langle j,k,l\rangle_{\alpha\beta}$ is shown in Fig.~\ref{fig1} (a).
{Perturbatively, the coefficient, $\kappa\propto\frac{h_xh_yh_z}{J_{\rm K}^2}$; a non-integrable interaction may  effectively enhance it further~\cite{PhysRevB.99.224409}. We rather
adopt Eq.~(\ref{eq:Hamiltonian}) as an effective model of CSL phase to compare with, e.g. the field-induced non-magnetic state of $\alpha$-RuCl$_3$, and consider $\kappa$ as a parameter to control its excitation gap.
We choose the Kitaev coupling, $J_{\rm K}$ as unit of energy, whose magnitude is roughly estimated as $J_{\rm K}\sim100$K$\sim1$meV, from neutron~\cite{BanerjeeNatMat} and Raman scattering on $\alpha$-RuCl$_3$~\cite{sandilands2015scattering}.}
Throughout the paper, we set $\kappa=0.1$, and adopt the system size, $N=16\times16\times2=512$ sites.

%---------%
\begin{figure*}[t]
\begin{center}
\includegraphics[width=\linewidth]{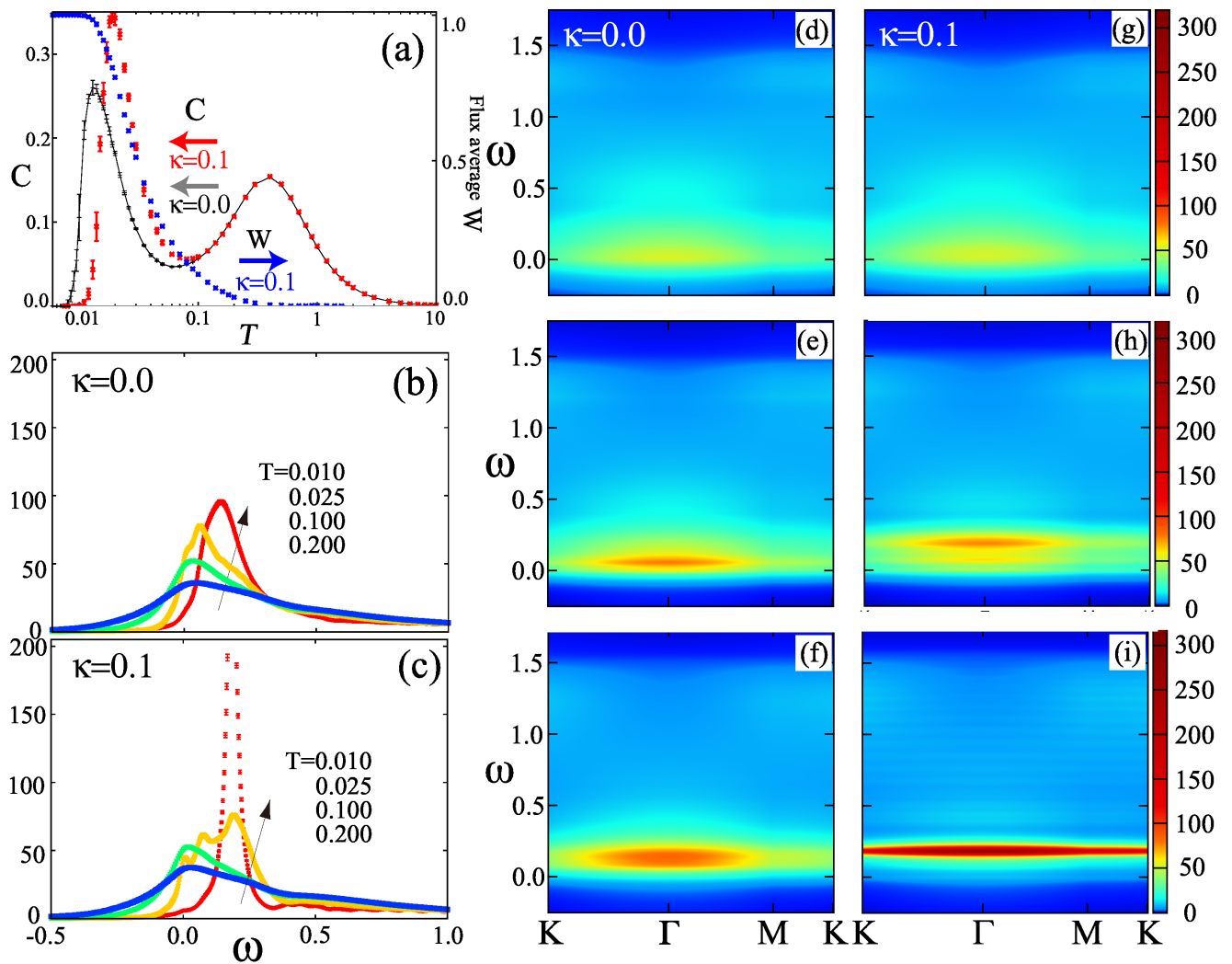}
\vspace{-10pt}
\caption{(Color online) (a) The specific heat and flux average at $\kappa=0.1$. The specific heat at $\kappa=0.0$ is also shown in black as a reference. (b)(c) The temperature dependence of $\mathcal{S}(\Gamma,\omega)$ at (b) $\kappa=0.0$ and (c) $\kappa=0.1$. (d)-(i) $\mathcal{S}({\mathbf q},\omega)$ on a symmetry line of Brillouin Zone for (d)-(f) $\kappa=0.0$ and (g)-(i) $\kappa=0.1$.
From the top, (d) (g) $T=0.1$, (e) (h) $T=0.025$, (f) (i) $T=0.012$.}
\label{fig2}
\end{center}
\end{figure*}
%---------% 

We rewrite the Hamiltonian in terms of Majorana operators ($c_j$),
\begin{eqnarray}
\mathcal{H}=\frac{i}{4}A_{ij}c_ic_j\equiv\frac{i}{4}\sum_{\langle i,j\rangle_{\alpha}}u_{ij}^{\alpha}c_ic_j + \frac{i}{4}\kappa\sum_{\langle j,k,l\rangle_{\alpha\beta}}u_{jl}^{\alpha}u_{kl}^{\beta}c_jc_k,
\label{eq:MajoranaHamiltonian}
\end{eqnarray}
{where the $u_{ij}^{\alpha}$ make up the conserved $\mathbb{Z}_2$ fluxes, $W_p$.}

In each configuration of $\{W_p\}$, we diagonalize the Hamiltonian matrix, $A$, to obtain $N/2$ pairs of eigenvalues, $(\varepsilon_m, -\varepsilon_m)$ with $\varepsilon_m>0$. Only the half of the eigenvalues are physical. We use the positive half of the eigenvalue spectrum to write the Hamiltonian in a diagonal form: $\mathcal{H}=\sum_{m=0}^{N/2-1}\varepsilon_m(2\gamma_m^{\dag}\gamma_m-1)$, where $\gamma_m$ is the fermionic eigenmode corresponding to $\varepsilon_m$. 
This form suggests that the system energy, $E$, can be written as the sum of the fermionic zero-point energy, $E_{\rm ZP}\equiv-\sum_m\varepsilon_m$, and the energy of excited fermions,
$ 
E = E_{\rm ZP} + \sum_{m:\gamma_m^{\dag}\gamma_m=1}2\varepsilon_m.
$ 
The flux free sector, $\{W_p=+1|\ \forall p\}$ gives the lowest $E_{\rm ZP}$, which corresponds to the ground state energy, $E_{\rm GS}$.
In this respect, the hexagon $p$ carrying $W_p=-1$ can be regarded as an excitation, the vison.

{\it Method:} To access the physical quantities at finite temperatures, we resort to the classical Monte Carlo simulation, by sampling $\{W_p=\pm1\}$.
In particular, we focus on the spin correlation function, $S^{\alpha}_{jj'}(\omega)\equiv\int_{-\infty}^{\infty}\frac{dt}{2\pi}\langle S_j^{\alpha}(t)S_{j'}^{\alpha}(0)\rangle e^{i(\omega+i\delta)t}$, where the real-time correlation function is obtained as,
\begin{widetext}
\begin{align}
&\langle S_j^{\alpha}(t)S_{j'}^{\alpha}(0)\rangle = \frac{1}{2Z}\sum_{\{W_p\}}\Bigl(\sqrt{{\rm det}(1 + e^{-(\beta-it)\cdot iA}e^{-it\cdot iA'})} \Bigl[\frac{1}{1 + e^{-(\beta-it)\cdot iA}e^{-it\cdot iA'}}e^{-(\beta-it)\cdot iA}\Bigr]_{j'j}\nonumber\\
&\hspace{0.2cm}-(-1)^{F_{\rm ph}}\sqrt{{\rm det}(1 - e^{-(\beta-it)\cdot iA}e^{-it\cdot iA'})} \Bigl[\frac{1}{1 - e^{-(\beta-it)\cdot iA}e^{-it\cdot iA^{(j)}}}e^{-(\beta-it)\cdot iA}\Bigr]_{j'j}\Bigr)(\delta_{j'j} - iu_{jj'}^{\alpha}\delta_{j'j_{\alpha}}).
\label{analyticalformula}
\end{align}
\end{widetext}
{Here, $A$ and $A'$ represent the Hamiltonian matrix in Eq.~(\ref{eq:MajoranaHamiltonian}), before and after the operation of $S_{j'}^{\alpha}(0)$ [Fig.~\ref{fig1} (b)]. $(-1)^{F_{\rm ph}}$ is a physical fermion parity, and $Z$ is the partition function.} For {more detail} of this equation, see Ref.~\cite{PhysRevB.98.220404}. The spin correlation is finite only up to a nearest-neighbor distance~\cite{PhysRevLett.98.247201} due to the property that a spin flip by operation $S_i^{\alpha}$ reverses a pair of $\mathbb{Z}_2$ fluxes on both sides of $\alpha$-bond extending from site $i$ [Fig.~\ref{fig1} (b)].
The dynamical magnetic structure factor is defined as
\begin{eqnarray}
\mathcal{S}({\mathbf q},\omega)=\frac{1}{N}\sum_{\alpha=x,y,z}\sum_{j,j'}e^{i{\mathbf q}\cdot({\mathbf r}_j-{\mathbf r}_{j'})}S^{\alpha}_{jj'}(\omega),
\end{eqnarray}
which is a relevant quantity to the inelastic neutron scattering experiments.

{\it Results.} To establish the gross features of the magnetic response,  we first show the specific heat, $C$ [Fig.~\ref{fig2} (a)], for $\kappa=0.1$ compared to $\kappa=0$.
Both show a characteristic 
two-peak structure~\cite{nasu2015thermal}. The first broad `Schottky' peak is on a scale of the fermion bandwidth $\sim J_K$; while the lower peak is associated with the visons, as evidenced by the concomitant appearance of a non-zero flux average, $W=\frac{2}{N}\sum_pW_p$: the cost
of inserting a vison pair into the flux-free state, $\Delta_{\rm V}$, which arises due to their difference in fermionic zero-point 
energies, $E_{\rm ZP}$, is a proxy for that energy scale. 
The specific heat is insensitive to the magnetic field at higher temperatures, while the vison peak is shifted upwards in energy {suggesting that $\Delta_{\rm V}$ increases with magnetic field.}

We now turn to the detailed analysis of the dynamical structure factor. 
$\mathcal{S}({\mathbf q}, \omega)$ is shown in Fig.~\ref{fig2} for $\kappa=0$ and $\kappa=0.1$ (middle and right rows) for (top to bottom) $T=0.1$, $0.025$, and $0.012$.

At the higher $T=0.1$, $\mathcal{S}({\mathbf q}, \omega)$ barely differs between the two cases, as was the case in the specific heat: both show broad intensity around $\omega=0$ and a weak continuous signal up to $\omega\sim1.5$, consistent with previous reports~\cite{yoshitake2016fractional}. 
For lower $T=0.025$, clear differences emerge. Whereas for $\kappa=0.0$, $\mathcal{S}({\mathbf q}, \omega)$, is  dominated by a broad peak around $\omega=0$, for $\kappa=0.1$, the dominant peak is found at finite energy, $\omega\sim0.2$. Moreover, this is accompanied by a multiplet of peaks at lower $\omega$.
At yet lower $T=0.012$, the broad peak for $\kappa=0.0$ persists, shifting upwards towards the vison gap $\Delta_{\rm V}$, whereas that 
for  $\kappa=0.1$ sharpens into a resonant peak at a higher energy. 

To better analyse the $T$-dependence in more detail, we plot  $\mathcal{S}(\Gamma, \omega)$ at the 
 $\Gamma$ point, ${\mathbf q}=0$.  
This again shows, for $\kappa=0.1$, a broad peak centred around $\omega=0$ at high temperature. Upon decreasing $T$, this zero-energy peak starts to diminish, and a sharp peak quickly evolves at finite energy, $\omega\sim0.2$. In the transient regime, $\mathcal{S}(\Gamma, \omega)$ shows a complicated structure, composed of three resonant peaks [Fig.~\ref{fig2} (c)]. 

%---------%
\begin{figure}[h]
\begin{center}
\includegraphics[width=0.99\linewidth]{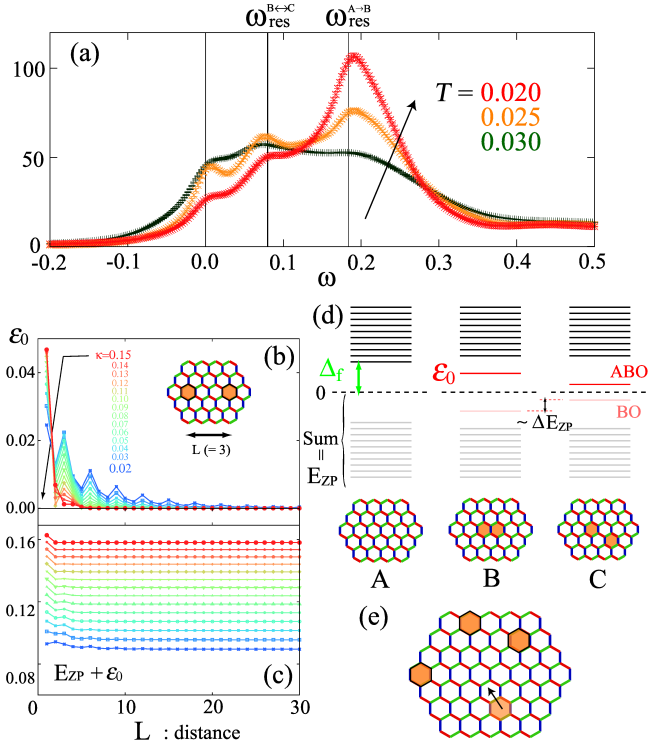}
%\vspace{-10pt}
\caption{(Color online) (a) $\mathcal{S}(\Gamma, \omega)$ in the transient regime, $T=0.030, 0.025, 0.020$. The three peaks in the spectrum are fitted with $\omega_{\rm res}=0, \omega^{{\rm B}\leftrightarrow{\rm C}}_{\rm res}$ and $\omega^{{\rm A}\rightarrow{\rm B}}_{\rm res}$, which are shown with vertical lines. (b) The bonding energy, $\varepsilon_0$ and (c) $E_{\rm ZP} + \varepsilon_0$ for each separation of vison pair, $L$, evaluated in a system of $N=64\times64\times2=8192$. (d) Schematic fermionic energy levels for the three flux configurations, A, B and C. Bonding (BO) and anti-bonding (ABO) orbitals are highlighted in red. (e) Schematic figure of the vison shift in a dilute vison regime, which gives rise to the zero-energy broad peak in (a).}
\label{fig3}
\end{center}
\end{figure}
%---------% 

We next argue that these features reflect the properties, discussed next, of the visons, and in particular, of the Majorana mode that accompanies them~\cite{PhysRevLett.86.268}. 
The fractionalisation of the spin degree of freedom tranlates into the observation that each single spin flip toggles the occupancy of one fermion mode, $m$, as well as the values of the fluxes in a pair of hexagons adjacent to the spin. 
The resulting resonance energy of process $i\to f$ is written as 
\begin{eqnarray}
\omega^{i\to f}_{\rm res} = 2\varepsilon^{(f)}_m + (E_{\rm ZP}^{(f)} - E_{\rm ZP}^{(i)}),
\label{eq:resonant_energy}
\end{eqnarray}
where $\varepsilon_m$ is the $m$-th fermionic level, and $E_{\rm ZP}$ is the fermionic zero-point energy, as defined above.

The crucial piece of physics at work now is the interaction between the modes residing on the visons. These form an anti-bonding orbital (ABO) and a bonding orbital (BO) at energies $\pm\varepsilon_0$, respectively [Fig.~\ref{fig3} (d): center], the energy splitting of which decreases, and vanishes exponentially, with the vison separation, $L$, [Fig.~\ref{fig3} (b)]. These `Majorana zero modes' reside in the gap, $\Delta_{\rm f}=\frac{3\sqrt{3}}{4}\kappa$ [Fig.~\ref{fig3} (d): left] of the fermionic spectrum. 

We now find that $E_{\rm ZP}+\varepsilon_0$ is only very weakly dependent on separation $L$ as long as $\kappa$ is sufficiently large [Fig.~\ref{fig3} (c)]; this means that the change of zero-point energy is dominated by the shift of, i.e.\ the interaction between, the vison Majorana modes, with the continuum levels contributing little difference, Fig.~\ref{fig3} (d).
Similar Majorana-mode interactions appear in the low-energy description of chiral superconductors~\cite{PhysRevB.92.134519,PhysRevB.94.060507} and non-Abelian fractional Hall systems~\cite{PhysRevB.85.161301}.

This leads to a reciprocal relation for the resonance frequency, associated with the MZM:
\begin{eqnarray}
2\varepsilon^{(f)}_0 + (E_{\rm ZP}^{(f)} - E_{\rm ZP}^{(i)}) = 2\varepsilon^{(i)}_0 + (E_{\rm ZP}^{(i)} - E_{\rm ZP}^{(f)}),
\label{eq:reciprocal}
\end{eqnarray}
i.e., a change between two flux configurations corresponds to the same energy in both directions!

We now use these insights to discuss the three peaks of $\mathcal{S}(\Gamma, \omega)$ at intermediate $T=0.025$ in Fig.~\ref{fig3} (a) in turn. 
The lowest of the three peaks  corresponds 
to shifting an  essentially isolated vison ($\varepsilon=0$, hence $\omega=0$) from one plaquette to the next, [Fig~\ref{fig3} (e)]. The intensity of this peak decreases at low $T$ with the number of thermally excited of visons, Fig~\ref{fig3} (a). 

By contrast, the highest-energy peak involves the pair-creation of visons, ${\rm A}\to{\rm B}$ in Fig.~\ref{fig3} (d). 
Assuming only a pair of neighboring visons exist, we obtain $E^{\rm B}_{\rm ZP}-E^{\rm A}_{\rm ZP}=0.095654$ and $\varepsilon^{\rm B}_0=0.043986$, which result in $\omega^{A\to B}_{\rm res}=0.183626$.
This value well fits the position of peak [Fig.~\ref{fig3} (a)].
This peak quickly evolves into the sharp resonant peak at low $T$, a delta-function at $T=0$~\cite{PhysRevB.92.115127}. 

The intermediate peak in turn directly reflects MZM-mediated attractive interaction between the visons. The relevant process here is the dissociation of a neighbouring vison pair,
${\rm B}\leftrightarrow{\rm C}$ in Fig.~\ref{fig3} (d): $E^{\rm C}_{\rm ZP}-E^{\rm B}_{\rm ZP}=0.0078212$ and $\varepsilon^{\rm C}_0=0.035740$ implies $\omega^{B\to C}_{\rm res}=0.0793012$. The energy of its reciprocal {process} is nearly 
equal $\omega^{C\to B}_{\rm res}=0.0801508$, due to the reciprocal relation, Eq.~(\ref{eq:reciprocal}).
We note that pair formation of visons was recently discussed, associated with the low-temperature resonant peak observed in Raman scattering~\cite{wulferding2019magnon}.

The three peaks vary differently upon changing temperature and magnetic field. Like the lowest peak, the middle one requires thermally excited visons, and its intensity hence decreases with $T$. 
By contrast, the third is not activated as it creates a vison pair from the vacuum, at a cost 
of the vison gap, $\sim\Delta_{\rm v}$, which increases with $\kappa$, resulting in the peak moving to higher $\omega$ with increasing $\kappa$. 

{\it Experiments:}
Finally, let us give a discussion on existing experiments.
Inelastic neutron scattering experiments found a broad peak at $\sim3$ meV in the field-induced paramagnetic region~\cite{Banerjeefield}.
It is tempting to associate it with the resonance peak due to vison pair creation [Fig.~\ref{fig2} (i)], as pointed out by the authors.
The major inconsistency with our analysis is the momentum dependence. 
In our analyses, $\mathcal{S}({\mathbf q}, \omega)$ is almost flat in the entire Brillouin zone [Fig.~\ref{fig2} (i)].
However, the observed peak is around the $\Gamma$ point, while the M point does not show substantial intensity.
This inconsistency may be resolved by considering additional Heisenberg and $\Gamma$ ($\Gamma'$) interactions not included in our analysis.
These interaction induce dynamics of the visons by violating the conserved nature of fluxes, which may alter the flat momentum dependence of $\mathcal{S}({\mathbf q}, \omega)$.

Raman scattering experiments find a quick growth of a resonant peak around 5 meV upon lowering $T$ to 2 K in the field-induced paramagnetic region~\cite{wulferding2019magnon}.
This was attributed to the formation of vison pair, corresponding to our central peak obtained in the transient regime [Fig.~\ref{fig3} (a)].
Considering the continuing growth of the observed peak as $T$ is lowered further, it however might invoke the pair-creation process, possibly assisted by $\Gamma$ ($\Gamma'$)-type interactions. These interactions give rise to a vison pair of type C in Fig.~\ref{fig3} (d), which will lead to a similar peak growth with the highest-energy peak in Fig.~\ref{fig3} (a),
and will continue to grow as lowering temperatures.

{\it Discussion}:
{How will the non-integrable interactions affect the resonant peaks?
These interactions endow the visons with kinetic energy, and turn their thermal assembly into a `vison metal'. The zero-energy resonant peak will be  transformed into `Drude peak', and it will accordingly stay at zero energy. We expect the higher two peaks to also persist as long as the bonding energy of visons dominates over their kinetic energy.}

{Vison dynamics will in turn affect the momentum structure of the resonant peaks. In the CSL phase, visons are expected to behave as Ising anyons.
It is thus interesting to clarify how their statistical property affects---and inversely how we can extract information from---the momentum dependence of the dynamical structure factor. The answer to this question opens an avenue to the long-awaited observation of braiding of non-Abelian anyons in a magnet. We would like to leave this interesting problem for future work.}

This work was supported by JSPS KAKENHI (Nos. JP15H05852, JP15K21717 and JP16H04026), MEXT, Japan, and by the Deutsche Forschungsgemeinschaft  under grants SFB 1143 (project-id 247310070) and the cluster of excellence ct.qmat (EXC 2147, project-id 39085490). We thank S. Nagler and D. Wulferding for useful information and discussions, and J. T. Chalker, J. Knolle and D. L. Kovrizhin for collaborations on related material.
\bibliographystyle{apsrev4-1}
\bibliography{ChiralSLneutron}

\end{document}